\documentclass[aps,twocolumn,showpacs]{revtex4}
\usepackage{graphicx}

\begin{document}

\title{Radiative Phase Transitions and Casmir Effect Instabilities}

\author{S. Sivasubramanian, A. Widom and Y.N. Srivastava$\dagger $}
\affiliation{Physics Department, Northeastern University, Boston MA USA\\
$\dagger $Physics Department \& INFN, University of Perugia, Perugia IT}

\begin{abstract}
Molecular quantum electrodynamics leads to photon frequency
shifts and thus to changes in condensed matter free energies often called
the Casimir effect. Strong quantum electrodynamic coupling between
radiation and molecular motions can lead to an instability beyond which
one or more photon oscillators undergo a displacement phase transition.
The phase boundary of the transition can be located by a Casimir free
energy instability.
\end{abstract}

\maketitle
\section{Introduction}
\label{intro}
The interaction between electromagnetic field and the molecules of
condensed matter shifts the frequency of the photon oscillators
which changes the free energy of the condensed matter system. This
free energy shift is known as the Casimir effect\cite{Casimir57}. From a quantum field
theoretical viewpoint, the Casimir free energy is the contribution
due to one photon loop Feynman diagrams\cite{Abrikosov75}. Originally, Casimir intended
to describe either photon oscillator induced forces between a few atoms
or between fixed perfect conductors. Presently Casimir effects include
the dielectric and paramagnetic materials as well as irreversible
effects in moving conductors\cite{Sassaroli00,Moore70,Fulling76,Dodonov96}.
The literature is quite extensive but several
excellent reviews are available\cite{Lamoreaux97,Bordag01,Milton01}.
Our purpose is to point out that if the coupling between the photon
oscillators and the molecules is
sufficiently strong, then the photon oscillators undergo a
displacement into a coherent electromagnetic radiation state. While the
single photon loops do not describe the nature of the radiation
coherence, the phase boundary between the normal radiation state and
the coherent radiation state can be located as a free energy instability.
As the thermodynamic equation of state boundary of the normal photon
oscillator regime is approached, one or more of the photon oscillator
frequencies tends toward zero. Such zero frequency modes lead to a Casimir
free energy signature of the forthcoming radiation transition.

In Sec.\ref{phase_shift}, the connection between the change in the photon
density of states free energy and the photon scattering phase shift is reviewed.
The Casimir free energy then follows by directly superimposing the Planck
free energies of the shifted oscillators. In Sec.\ref{propagator}, the free
energy is expressed directly in terms of the (gauge invariant) electric field
photon propagator \begin{math} {\sf G} \end{math} and the dielectric
susceptibility \begin{math} {\bf \chi } \end{math}. The Casimir instability
is explored in Sec.\ref{Polarization} (eventually in terms of the dielectric
response function \begin{math}  \varepsilon  \end{math}). The phase
regime of the Casimir photon oscillator stability is shown to be
identical to the phase regime of the stable Clausius-Mossotti
theory\cite{Mossotti50,Clausius97}. The general physical principles
involved are discussed in the concluding Sec.\ref{physics}.

\section{Photon Scattering Operator}
\label{phase_shift}
Consider a condensed matter object which can elastically scatter
photons. Mathematically the scattering of a photon having frequency
\begin{math} \omega  \end{math} would be described by a unitary
scattering operator \begin{math} S(\omega ) \end{math} whose
eigenvalues determine a set of phase shifts
\begin{math} \{\delta_\nu (\omega )\}  \end{math} via
\begin{equation}
S(\omega^2 )\left|\nu \right>=e^{2i\delta_\nu (\omega)}\left|\nu \right>.
\label{scatter1}
\end{equation}
The Wigner time delay\cite{Wigner55} \begin{math} \tau_\nu (\omega ) \end{math} in
a photon scattering channel \begin{math} \nu \end{math}
represents the extra amount of time the photon remains in the
neighborhood of the target due to the photon-target interaction.
It is well known to obey
\begin{equation}
\tau_\nu (\omega )=
2\frac{d\delta_\nu (\omega )}{d\omega }\ .
\label{scatter2}
\end{equation}
If one sums the time delays over all the channels, then one may
compute the extra number \begin{math} dN(\omega ) \end{math}
of photon states in a bandwidth \begin{math} d\omega  \end{math}
due the target-photon interaction
\begin{equation}
dN(\omega )=\frac{1}{2\pi }\sum_\nu \tau_\nu (\omega )d\omega
=\frac{1}{\pi }\sum_\nu d\delta_\nu (\omega ).
\label{scatter3}
\end{equation}
The Casimir contribution to the condensed matter object free energy
may be written in terms of the free energy
\begin{math} f(\omega,T) \end{math}
of a simple harmonic oscillator; i.e.
\begin{eqnarray}
f(\omega,T)&=&
-k_BT\ln\left\{\sum_{N=0}^\infty e^{-(N+1/2)\hbar\omega /k_BT}\right\}
\nonumber \\
&=& k_BT\ln
\left\{2\sinh\left(\frac{\hbar \omega}{2k_BT}\right)\right\},
\nonumber \\
\frac{\partial f(\omega ,T)}{\partial\omega }&=&
\frac{\hbar }{2}\coth\left(\frac{\hbar \omega }{2k_BT}\right)
=k_BT\sum_{n=-\infty}^\infty \frac{\omega }{\omega^2+\omega_n^2}\ ,
\nonumber \\
\omega_n &=& \left(\frac{2\pi k_BT}{\hbar }\right)n,
\ \ \ n=0,\pm 1,\pm 2,ldots \ .
\label{scatter4}
\end{eqnarray}
The Casimir free energy is found by simply summing oscillator
free energies over all the added photon modes as dictated by
Eqs.(\ref{scatter3}) and (\ref{scatter4}); It is
\begin{eqnarray}
\Delta F &=& \int_0^\infty f(\omega ,T)dN(\omega)=-\int_0^\infty
\frac{\partial f(\omega ,T)}{\partial \omega }N(\omega)d\omega ,
\nonumber \\
\Delta F &=& -k_BT \sum_{n=-\infty}^\infty \int_0^\infty
\frac{\omega N(\omega )d\omega }{\omega^2+\omega_n^2}
\nonumber \\
&=& -\frac{k_BT}{\pi}\sum_{n=-\infty}^\infty \int_0^\infty
\frac{\omega \{\sum_\nu \delta_\nu (\omega )\}d\omega }{\omega^2+\omega_n^2} .
\label{scatter5}
\end{eqnarray}
Now consider the analytic continuation from real frequency
\begin{math} \omega  \end{math} to the complex frequency
\begin{math} \zeta  \end{math} in the upper half plane
\begin{equation}
\zeta =\omega +is \ \ \ {\rm where}\ \ \ s\equiv{\Im}m\zeta \ge 0.
\label{scatter6}
\end{equation}
The analytically continued scattering operator obeys the dispersion relation
\begin{eqnarray}
\ln \det S(\zeta ^2)&=&\frac{2}{\pi }\int_0^\infty
\frac{\omega {\Im}m\ln \det S(\omega ^2+i0^+)d\omega }{\omega^2-\zeta ^2}\ ,
\nonumber \\
\ln \det S(\zeta ^2)&=&\frac{4}{\pi }\int_0^\infty
\frac{\omega \{\sum_\nu \delta_\nu (\omega )\}d\omega }{\omega^2-\zeta ^2}\ .
\label{scatter7}
\end{eqnarray}
Our final expression for the Casimir free energy in terms of the analytically
continued scattering operator follows from Eqs.(\ref{scatter5}) and
(\ref{scatter7}) to be
\begin{equation}
\Delta F =-\frac{k_BT}{4}\sum_{n=-\infty }^\infty \ln \det S(-\omega_n ^2).
\label{scatter8}
\end{equation}
Let us now consider the more general case wherein the scattering operator
may also describe inelastic processes such as photon absorption by the
condensed matter sample. The eigenvalue problem,
\begin{math}
S\left|\nu \right>=e^{2i\Delta_\nu }\left|\nu \right>
\end{math}, now yields complex phase shifts with
\begin{math}{\Im }m\ \Delta_\nu \ge 0 \end{math}.
However, the Casimir free energy Eq.(\ref{scatter8}) remains
in tact. This is most easily understood in terms of the photon
propagator.
\section{Photon Propagators}
\label{propagator}

Starting from Maxwell's equations
\begin{eqnarray}
curl{\bf E}&=&-\frac{1}{c}\left(\frac{\partial {\bf B}}{\partial t}\right),
\nonumber \\
curl{\bf B}&=&\frac{1}{c}\left(\frac{\partial {\bf E}}{\partial t}\right)
+\frac{4\pi }{c}{\bf J},
\nonumber \\
div{\bf E}&=&4\pi \rho\ \ {\rm and}\ \ div{\bf B}=0,
\label{propagator2}
\end{eqnarray}
one finds that
\begin{equation}
\left\{\frac{1}{c^2}
\left(\frac{\partial }{\partial t}\right)^2-\Delta
\right\}{\bf E}=-4\pi
\left\{ \frac{1}{c^2}\frac{\partial {\bf J}}{\partial t}+
{\bf grad }{\rho } \right\}.
\label{propagator3}
\end{equation}
For any locally conserved charge
\begin{math} (\partial \rho /\partial t)+div{\bf J}=0  \end{math},
there exists a polarization \begin{math} {\bf P} \end{math} such that
\begin{equation}
\rho =-div{\bf P}\ \ {\rm and}\ \ {\bf J}=\frac{\partial {\bf P}}{\partial t}\ .
\label{propagator4}
\end{equation}
If we invoke the retarded tensor propagator
\begin{equation}
{\sf G}({\bf r},\zeta )=\left\{\left(\frac{\zeta }{c}\right)^2{\sf 1}
+{\bf \nabla \nabla }\right\}\frac{e^{i\zeta r/c}}{r}\ \
{\rm for}\ \ {\Im}m\ \zeta >0,
\label{propagator5}
\end{equation}
then Eqs.(\ref{propagator3}) and (\ref{propagator4}) are formally solved by
\begin{eqnarray}
{\bf E}({\bf r},t)&=&{\bf E}_{in}({\bf r},t)+
\nonumber \\
&\ &\int {\sf G}
\left({\bf r}-{\bf r}^\prime ,\zeta=i\frac{\partial }{\partial t}\right)
\cdot {\bf P}({\bf r}^\prime ,t)d^3{\bf r}^\prime ,
\label{propagator6}
\end{eqnarray}
wherein the incoming photon is described by an electric field
obeying the vacuum wave equation
\begin{equation}
\left\{\frac{1}{c^2}
\left(\frac{\partial }{\partial t}\right)^2-\Delta
\right\}{\bf E}_{in}=0,
\label{propagator7}
\end{equation}
and the outgoing photon has been scattered by the polarization
\begin{math} {\bf P} \end{math}. The use of the formal operator
replacement \begin{math} \zeta =i(\partial /\partial t) \end{math}
may be illustrated as follows: The polarization response in the condensed
matter target may be related to the electric field by employing a
retarded non-local response function
\begin{eqnarray}
{\bf P}({\bf r},t)&=&\int_0^\infty  \int {\sf K}({\bf r},{\bf r}^\prime ,s)
\cdot {\bf E}({\bf r}^\prime ,t-s)d^3{\bf r}^\prime ds,
\nonumber \\
{\bf P}({\bf r},t)&=& \int \left[\int_0^\infty{\sf K}({\bf r},{\bf r}^\prime ,s)
e^{-s(\partial /\partial t)}ds\right]
\cdot {\bf E}({\bf r}^\prime ,t)d^3{\bf r}^\prime ,
\nonumber \\
{\bf P}({\bf r},t)&=&\int {\bf \chi}
\left({\bf r},{\bf r}^\prime ,\zeta =i\frac{\partial }{\partial t}\right)
\cdot {\bf E}({\bf r}^\prime ,t)d^3{\bf r}^\prime ,
\label{propagator8}
\end{eqnarray}
with the complex frequency dependent susceptibility defined as
\begin{equation}
{\bf \chi}({\bf r},{\bf r}^\prime ,\zeta )=
\int_0^\infty{\sf K}({\bf r},{\bf r}^\prime ,s) e^{i\zeta s}ds .
\label{propagator9}
\end{equation}
The scattering equation follows from Eqs.(\ref{propagator6}) and
(\ref{propagator8}) to be
\begin{eqnarray}
{\bf E}({\bf r},t) &=& {\bf E}_{in}({\bf r},t)+ \int \int { \sf G}
\left({\bf r}-{\bf r}^\prime ,\zeta = i\frac{\partial }{\partial t}\right)
\nonumber \\
\cdot&{\bf \chi}&
\left({\bf r}^\prime ,{\bf r}^{\prime \prime },
\zeta =i\frac{\partial }{\partial t}\right)
\cdot {\bf E}({\bf r}^{\prime \prime } ,t)d^3{\bf r}^\prime
d^3{\bf r}^{\prime \prime}.
\label{propagator10}
\end{eqnarray}
The Fredholm-Jost determinant\cite{Newton82,Jost51}
\begin{math} J(\zeta ) \end{math} and its
logarithmic determinant \begin{math} \Phi (\zeta ) \end{math} for the
scattering integral Eq.(\ref{propagator10}) are defined as
\begin{eqnarray}
J(\zeta )&=&\det \left\{1-G(\zeta )\chi (\zeta ) \right\},
\nonumber \\
\Phi (\zeta )&=&-\ln J(\zeta )
=-Tr\ln\left\{1-G(\zeta )\chi (\zeta ) \right\}.
\label{propagator11}
\end{eqnarray}
Eq.(\ref{propagator11}) is related to the determinant of the scattering
operator according to
\begin{equation}
\det S(\zeta^2 )=-\left[\frac{J(\zeta )}{J(-\zeta )}\right] .
\label{propagator12}
\end{equation}
\begin{figure}[bp]
\scalebox {0.7}{\includegraphics{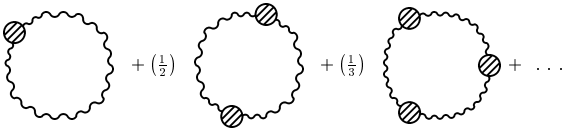}}
\caption{Shown are the one loop Feynman diagrams contributing to the Casimir
free energy in Eq.(\ref{propagator14}). For each propagator wavy line
one inserts ${\sf G}_{ij}({\bf r}_i,{\bf r}_j,\zeta )$ in accordance with
Eq.(\ref{propagator5}). For each ``self energy part'' one inserts
$\chi_{kl}({\bf r}_k,{\bf r}_l,\zeta )$ in accordance with
Eqs.(\ref{propagator8}) and (\ref{propagator9}). The trace ``$Tr$''
includes sums over polarization indices and integrals over space.}
\label{fig1}
\end{figure}
From Eqs.(\ref{scatter8}), (\ref{propagator11}) and (\ref{propagator12})
it then follows that
\begin{eqnarray}
\Delta F =-\frac{k_BT}{2}\sum_{n=-\infty }^\infty \Phi (i|\omega_n |),
\nonumber \\
\Phi (\zeta )=
\sum_{k=1}^\infty \frac{1}{k}Tr \left\{G(\zeta )\chi (\zeta )\right\}^k .
\label{propagator14}
\end{eqnarray}
Eqs.(\ref{propagator14}) yield the ``one loop'' contributions to the
free energy which constitutes the Casimir effect. The Feynman diagrams are
shown in Fig.\ref{fig1}. If one employs the eigenvalue equation for
a photon state \begin{math} \left|\nu \right>  \end{math},
\begin{equation}
G(\zeta )\chi (\zeta )\left|\nu \right>=\eta_\nu (\zeta )\left|\nu \right>,
\label{propagator15}
\end{equation}
then
\begin{equation}
\Phi (\zeta )=\sum_\nu \sum_{k=1}^\infty \frac{\eta_\nu (\zeta )^k }{k}
=-\sum_\nu \ln \Big(1-\eta_\nu (\zeta )\Big).
\label{propagator16}
\end{equation}
For the Casimir free energy \begin{math} \Delta F  \end{math}
in Eqs.(\ref{propagator14}) and (\ref{propagator16}) to be stable,
it is necessary for the eigenvalues to obey the condition
\begin{equation}
\eta_\nu (i|\omega_n| )<1 \ \ {\rm for\ all}\ \nu \ {\rm and }\ n, \ \ \
({\rm Stable}).
\label{propagator17}
\end{equation}
If any eigenvalue (say as a function of temperature) passes through
unity, then the Casimir free energy will have an imaginary part. The
transition rate per unit time for the system to make a transition
into a stable state is given by
\begin{math} \Gamma =-(2/\hbar ){\Im}m\ \Delta F \end{math}.
The Casimir one photon loop free energy in Fig.\ref{fig1} is
not sufficient for computing the free energy of the stable ordered
radiative phase. Nevertheless, by examining the eigenvalue
Eq.(\ref{propagator15}) equation for the mode, i.e. channel
\begin{math} \left|\nu \right>  \end{math}, which becomes unstable,
one may gain insights into the radiative coherent state which will
be stabilized. The phase diagram for the disordered to ordered
radiative state may also be computed. Let us illustrate
how this comes about.
\section{Polarization Instability}
\label{Polarization}

From Eqs.(\ref{propagator6}), (\ref{propagator8}) and (\ref{propagator15})
one may examine the electric field of a channel mode
\begin{math} \left|\nu \right>  \end{math} according to
\begin{eqnarray}
\int \int {\sf G}({\bf r}-{\bf r}^\prime ,\zeta)
&\cdot& {\bf \chi}({\bf r}^\prime ,{\bf r}^{\prime \prime },
\zeta )
\nonumber \\
&\cdot& {\bf E}_\nu ({\bf r}^{\prime \prime })d^3{\bf r}^\prime
d^3{\bf r}^{\prime \prime }=\eta_\nu (\zeta ){\bf E}_\nu ({\bf r}).
\label{polar1}
\end{eqnarray}
Eq.(\ref{polar1}) can be decomposed into two parts: (i) The mode
\begin{math} {\bf E}_\nu ({\bf r}) \end{math} produces a polarization
\begin{math} {\bf P}_\nu ({\bf r}) \end{math}. (ii) The
polarization \begin{math} {\bf P}_\nu ({\bf r}) \end{math} radiates the
electric field \begin{math} {\bf E}_\nu ({\bf r}) \end{math}.
\begin{eqnarray}
\int {\bf \chi}({\bf r},{\bf r}^\prime ,\zeta )\cdot
{\bf E}_\nu ({\bf r}^\prime )d^3{\bf r}^\prime &=& \eta_\nu (\zeta )
{\bf P}_\nu({\bf r}),
\nonumber \\
\int {\sf G}({\bf r}-{\bf r}^\prime ,\zeta )\cdot
{\bf P}_\nu ({\bf r}^\prime )d^3{\bf r}^\prime &=&{\bf E}_\nu({\bf r}).
\label{polar2}
\end{eqnarray}
If the eigenvalue \begin{math} \eta  \end{math} passes through unity,
then the field mode can be sustained self consistently indicating a
radiative instability. Otherwise, the coupling is two weak to maintain
the mode and the Casimir free energy contribution is stable.

To see what is involved, let us consider a translational invariant
fluid for which the bulk Casimir free energy per unit volume
\begin{math} \Delta f \end{math} is determined by electric susceptibility
\begin{equation}
{\bf \chi}({\bf r}-{\bf r}^\prime ,\zeta )=
\int {\bf \chi}_{\bf k}(\zeta )e^{i{\bf k}\cdot ({\bf r}-{\bf r}^\prime)}
\frac{d^3{\bf k}}{(2\pi )^3}\ .
\label{polar3}
\end{equation}
In detail, Eqs.(\ref{propagator5}),(\ref{propagator14}),
(\ref{propagator16}), (\ref{polar2}),
and (\ref{polar3}) imply
\begin{eqnarray}
{\bf E}({\bf r};{\bf k}) &=& {\bf E}_{\bf k}e^{i{\bf k\cdot r}},
\nonumber \\
{\sf G}_{\bf k}(\zeta ) &=&\int {\sf G}({\bf r},\zeta )
e^{-i{\bf k\cdot r}}d^3 {\bf r}
\nonumber \\
&=& -4\pi \left\{
\frac{{\bf kk}+(\zeta /c)^2{\bf 1}}{k^2-(\zeta /c)^2}\right\},
\nonumber \\
\eta_{\bf k}(\zeta ){\bf E}_{\bf k}&=&
{\sf G}_{\bf k}(\zeta )\cdot {\bf \chi}_{\bf k}(\zeta )\cdot {\bf E}_{\bf k},
\nonumber \\
\frac{2\Delta f}{k_BT} &=& -\sum_{n=-\infty }^\infty \int
\ln \Big(1-\eta_{\bf k}(i|\omega_n|)\Big)\frac{d^3{\bf k}}{(2\pi )^3}.
\label{polar4}
\end{eqnarray}
The {\em static} (\begin{math} \omega_n=0 \end{math}) contribution to the
free energy per unit volume in Eq.(\ref{polar4}) is then determined by
\begin{eqnarray}
\varepsilon ({\bf k}) &=& 1-\eta_{\bf k}(0)
\nonumber \\
\Delta f_{static}&=&-\frac{k_BT}{2}\int \ln
\varepsilon ({\bf k})\ \frac{d^3{\bf k}}{(2\pi )^3}\ ,
\nonumber \\
\varepsilon ({\bf k})&=&1+4\pi\left(
\frac{{\bf k}\cdot {\bf \chi }_{\bf k}(0)\cdot {\bf k}}{k^2}\right)
=1+4\pi\tilde{\chi }({\bf k}).
\label{polar5}
\end{eqnarray}

Let us examine Eqs.(\ref{propagator5}) and (\ref{polar2}) in the limit
of zero frequency; i.e.
\begin{eqnarray}
\lim_{\zeta \to 0} {\sf G}({\bf r}-{\bf r}^\prime ,\zeta )
&=& {\bf \nabla \nabla }\left(\frac{1}{|{\bf r}-{\bf r}^\prime |}\right)
\nonumber \\
\lim_{\zeta \to 0} {\sf G}({\bf r}-{\bf r}^\prime ,\zeta )
&=&{\sf D}({\bf r}-{\bf r}^\prime )
-\left(\frac{4\pi }{3}\right){\bf 1}\delta ({\bf r}-{\bf r}^\prime ),
\nonumber \\
{\sf D}({\bf r})&=&\left\{\frac{3{\bf r}{\bf r}-r^2{\bf 1}}{r^5}\right\} ,
\nonumber \\
{\bf E}_{\bf k}({\bf r})&=&
\int {\sf D}({\bf r}-{\bf r}^\prime )\cdot
{\bf P}_{\bf k}({\bf r}^\prime )d^3{\bf r}^\prime
\nonumber \\ &-&
\left(\frac{4\pi }{3}\right){\bf P}_{\bf k}({\bf r}).
\label{polar6}
\end{eqnarray}
A static polarization will produce a local dipolar electric field
\begin{math} {\bf F}({\bf r};{\bf k}) \end{math} via
\begin{eqnarray}
{\bf F}({\bf r};{\bf k}) &=&
\int {\sf D}({\bf r}-{\bf r}^\prime )\cdot
{\bf P}({\bf r}^\prime ;{\bf k})d^3{\bf r}^\prime ,
\nonumber \\
{\bf F}({\bf r};{\bf k}) &=& {\bf E}({\bf r};{\bf k})+
\left(\frac{4\pi }{3}\right){\bf P}({\bf r};{\bf k}).
\label{polar7}
\end{eqnarray}
For a fluid with a volume per molecule of \begin{math} v \end{math}
and a nonlocal molecular polarizability
\begin{math} \beta ({\bf k})\end{math}, defined by the local field
\begin{math} {\bf F}  \end{math},
\begin{equation}
v{\bf P}({\bf r};{\bf k})=\beta ({\bf k}){\bf F}({\bf r};{\bf k}),
\label{polar8}
\end{equation}
Eqs.(\ref{polar7}) and (\ref{polar8}) imply the dielectric response function
\begin{eqnarray}
{\bf D}({\bf r};{\bf k})&=&
{\bf E}({\bf r};{\bf k})+4\pi {\bf P}({\bf r};{\bf k})
=\varepsilon({\bf k}){\bf E}({\bf r};{\bf k}),
\nonumber \\
\varepsilon({\bf k}) &=& \frac{1+\{8\pi \beta ({\bf k})/3v\}}
{1-\{4\pi \beta ({\bf k})/3v\}}\ .
\label{polar9}
\end{eqnarray}

Note, in the limit of zero wave vector Eq.(\ref{polar9}) yields the usual
Clausius-Mossotti formula relating the static dielectric constant to
the molecular polarizability
\begin{eqnarray}
\lim_{|{\bf k}|\to 0}\beta ({\bf k})&=& \alpha \ \ \ {\rm (polarizability)},
\nonumber \\
\lim_{|{\bf k}|\to 0}\varepsilon({\bf k}) &=& \varepsilon
\ \ \ {\rm (dielectric\ constant)},
\nonumber \\
\varepsilon &=& \frac{3v+8\pi \alpha }
{3v-4\pi \alpha }\ .
\label{polar10}
\end{eqnarray}
The Clausius-Mossotti stability condition\cite{Siva02_1,Siva02_2},
\begin{equation}
3v > 4\pi \alpha \ \ \ ({\rm {stable\ normal\ phase}}),
\label{polar11}
\end{equation}
has then been shown to follow from the Casimir effect stability
(\begin{math} \eta<1 \end{math}).

\section{Conclusions}
\label{physics}

The quantum electrodynamic interaction between molecules and photons
gives rise to frequency Lamb shifts\cite{Lamb47} which in turn
contribute to a shift in the free energy of a condensed matter system.
This Casimir contribution to the free energy can become unstable if the
interaction coupling is too strong. The instability is the signature
to a phase transition into a state with coherent radiation. We have
exhibited such an instability within the above statistical thermodynamic
considerations. We conclude this work by exhibiting the instability
directly in terms of a collective Lamb shift.

For zero wave number (\begin{math} {\bf k}=0 \end{math}) but finite frequency
(\begin{math} \omega \ne 0 \end{math}), the dielectric response function of
Eq.(\ref{polar9}) reads
\begin{equation}
\varepsilon(\omega) = \frac{1+\{8\pi \beta (\omega)/3v\}}
{1-\{4\pi \beta (\omega)/3v\}}\ .
\label{physics1}
\end{equation}
If the molecular polarizability \begin{math} \beta(\omega ) \end{math}
can be described by a single excitation
frequency \begin{math} \omega_0=(\Delta E/\hbar )  \end{math} via
\begin{equation}
\beta(\omega) = \left(\frac{\omega_0^2}{\omega_0^2-\omega ^2}\right)\alpha ,
\label{physics2}
\end{equation}
then the dielectric response function
\begin{equation}
\varepsilon(\omega) = \frac{3v(\omega_0^2-\omega ^2)+8\pi \omega_0^2\alpha }
{3v(\omega_0^2-\omega ^2)-4\pi \omega_0^2\alpha }
\label{physics3}
\end{equation}
gives rise to a shifted frequency
\begin{equation}
\Omega_0^2=\omega_0^2\left\{1-\frac{4\pi \alpha }{3v}\right\}
\ \ \ \ \ \  ({\rm collective\ Lamb\ Shift}).
\label{physics4}
\end{equation}
The Eq.(\ref{polar11}) for the stable phase regime requires that
the collective Lamb shifted frequency \begin{math} \Omega_0 \end{math}
be real. A strong coupling imaginary frequency is a clear indicator of
an unstable radiation phase.

While the Casimir effect expressions are sufficient to derive the
{\em stable} photon oscillator regime in temperature and density,
the one photon loop approximation lacks the sensitivity to decide
the nature of the radiation ordered state. Ordered states of matter
are not easily understood from a perturbation theory viewpoint.

\end{document}